# Electric field driven switching of individual magnetic skyrmions

Pin-Jui Hsu, André Kubetzka, Aurore Finco, Niklas Romming,
Kirsten von Bergmann*, Roland Wiesendanger

*Department of Physics, University of Hamburg, 20355 Hamburg, Germany*

*Corresponding author. E-mail: kbergman@physnet.uni-hamburg.de

**Controlling magnetism with electric fields is a key challenge to develop future energy-efficient devices, however, the switching between inversion symmetric states, e.g. magnetization up and down as used in current technology, is not straightforward, since the electric field does not break time-reversal symmetry. Here, we demonstrate that local electric fields can be used to reversibly switch between a magnetic skyrmion and the ferromagnetic state. These two states are topologically inequivalent, and we find that the direction of an electric field directly determines the final state. This observation establishes the possibility to combine energy-efficient electric field writing with the recently envisaged skyrmion racetrack-type memories.**

Current magnetic information technology is mainly based on writing processes requiring either local magnetic fields or spin torques, which are both generated by currents and thus inherently imply large switching power. It has been demonstrated that magnetic properties at surfaces or interfaces can be altered upon the application of large electric fields (*1-5*). This has mostly been ascribed to changes in magnetocrystalline anisotropy due to spin-dependent surface screening and modifications of the band structure (*6-8*), changes in atom positions (*5,9,10*), or differences in hybridization with an adjacent oxide layer (*4,11*). Since the electric field does not break time-reversal symmetry, several workarounds have been proposed to toggle between bistable magnetic states with electric fields (*12,13*). Even a change of material composition due to electric fields has been presented as an alternative to switch between states with different magnetic properties (*14*).

This fundamental hurdle might be circumvented altogether by using magnetic skyrmions as information carriers instead of conventional bistable states. Magnetic skyrmions represent knots in the spin texture and thus are topologically distinct from the trivial ferromagnetic state (*15,16*). They can form in magnetic systems with broken inversion symmetry due to the Dzyaloshinskii-Moriya interaction (DMI), which is a consequence of spin-orbit coupling and favors an orthogonal spin configuration with a material-specific rotational sense. For a utilization of skyrmions in future spintronic devices it is indispensable to be able to reliably write and delete them individually. A local switching between the skyrmion and the ferromagnetic state has been demonstrated experimentally with vertically injected spin-polarized currents from a scanning tunneling microscope (STM) tip (*17*). However, non-collinear magnetic states are highly susceptible to electric currents, which may lead to a movement of magnetic skyrmions above a surprisingly low current threshold (*15,18,19*). While this is a benefit on one hand, as information can be transported easily through the material (*20,21*), in a write unit it is desirable to encode the information in a specific position, without a subsequent movement of the written magnetic bit.

In this work we show that an electric field, rather than a local spin current, can be used to reliably switch between a skyrmionic and the ferromagnetic state, thereby offering great potential for future energy-efficient skyrmion based memory and logic devices. Our model system is a three atomic layer thick epitaxial Fe film on an Ir(111) substrate and Figure 1a shows the topography of a typical sample. Due to the lattice mismatch, dislocation lines with a periodicity between 4 and 9 nm are observed in the Fe film, a manifestation of strain relief. Figure 1b shows the spin-resolved map of differential conductance (d$I$/d$U$) of the area indicated by the dashed box in 1a. It demonstrates that the magnetic ground state in zero magnetic field is a spin spiral, which propagates in the direction of the dislocation lines. In this area the spin spiral has a magnetic period of about 3.8 nm. Similar to the spin spiral state of the Fe double layer on the same substrate (*22*), the wavefronts exhibit a zigzag shape with the periodicity of the dislocation lines. This is due to a coupling of the magnetic state to the structure of the reconstructed Fe layer (see supplementary materials, Fig. S1). When an external magnetic field is applied, Fig. 1c, the magnetic state changes: at +2.5 T the bright areas have grown, and only individual magnetic objects are left. The smallest units exhibit a bean-like shape, reflecting the symmetry of the Fe atom arrangement in the layer.

To characterize the spin structure of the bean-shaped magnetic objects, we measure the in-plane magnetization components of all three possible rotational domains, Fig. 2a, i.e. on magnetic objects rotated by 120° with respect to each other, see spin-resolved d$I$/d$U$ maps in Fig. 2b-d. While the magnetic entities have a different appearance for each rotational domain, they all look identical within one rotational domain; for instance in Fig. 2c the objects are always imaged bright at the bottom and dark at the top. Combining this with the knowledge from Fig. 1c, i.e. that the out-of-plane magnetization component is antiparallel to the external magnetic field in the center and along the applied field between the magnetic objects, we can conclude that all of them have the same rotational sense. This is not surprising, as it is known that the DMI is strong at the Fe/Ir(111) interface (*23*). Figure 2e-g (top) show spin-resolved d$I$/d$U$ maps of one representative magnetic object from each rotational domain, rotated to have the dislocation lines run vertically. This is equivalent to imaging the same magnetic state with sensitivity to three different in-plane magnetization directions as indicated by the arrows. This allows for a reconstruction of the spin structure (*23*), and the result is shown in Fig. 2h. Simulations of SP-STM images (*24*) of this spin configuration, shown in Figs. 2e-g (bottom), agree well with the experimental data (top). While due to the underlying atom arrangement this magnetic object is not axially symmetric like magnetic skyrmions in an isotropic environment, see sketch in Fig. 2i, it has the same characteristics of unique rotational sense of the magnetization, which wraps the unit sphere, and thus it exhibits the topology of a skyrmion.

In view of future spintronic applications based on magnetic skyrmions, it is essential to be able to write and delete them, which we demonstrate for our system in the measurement series presented in Fig. 3: a negative voltage of -3 V, i.e. a tunnel current from sample to tip, deletes a skyrmion, and the opposite current direction with +3 V writes it. Since the bias polarity determines the direction of the switching process, we can rule out that non-directional effects like Joule heating or the finite measurement temperature play an important role; instead this finding is a hallmark of either spin-transfer torque due to the flow of spin-polarized currents (*17*), or a signature of electric field driven effects, both possible in an SP-STM setup, cf. Fig. 3 (top). To distinguish between these two mechanisms we use a non-magnetic W tip, for which the spin-transfer torque contribution vanishes. Due to the different electronic nature of non-collinear and ferromagnetic states, a skyrmion can also be detected with spin-averaging tips (*25*), as seen in the image series of Fig. 4a. Here, the ability to annihilate and create skyrmions with a non-magnetic electrode is demonstrated and suggests a decisive role of the electric field.

The electric field between an STM tip and the sample surface can be estimated via the ratio of applied sample bias voltage and tip-sample distance, $E = U/d$. Due to the shape of the tip the electric field derived in such a parallel plate model is an upper bound for the electric field at the surface (*26*). While $U$ and relative tip-sample distances are directly accessible in STM experiments we estimate the absolute distance $d$ by extrapolation (see Fig. S2). To determine the threshold voltage $U_t$ for switching the magnetic state, voltage sweeps with $d$ kept constant were performed. In such a measurement the switching of a skyrmion manifests itself in a jump in the tunnel current. We observe successful writing and deleting of the skyrmion nearly every time we perform the voltage sweep (see also Fig. S3). Note that the exact values of $U_t$ depend slightly on the used micro-tip and the distance between dislocation lines and thus only measurements with the same tip on the identical sample spot, as presented in Fig. 4a,b, can be compared quantitatively (see also Fig. S4). The threshold voltage to switch the right skyrmion in Fig. 4a was measured for different tip-sample distances, as well as in different magnetic fields, Fig. 4b. The linear behavior of $U_t$ with respect to the tip-sample distance at each magnetic field demonstrates that indeed a critical electric field $E_c$ is responsible for the switching of the magnetic state. From a linear fit of all data with the constraint of a joint crossing at zero bias we derive electric fields of $E_c = 1 - 6$ V/nm. Due to the STM tip geometry these values are upper bounds, meaning that the actual electric fields necessary for writing or deleting might be lower. We conclude that electric fields allow for controlled and reliable switching between topologically distinct magnetic states.

Different mechanisms for the electric field driven switching are possible. Given that the electric field influences the charge distribution in the surface region and possibly the interatomic distances (*1-10*), a modification of the relative strengths of the magnetic energies is expected. Considering that the formation of skyrmions in external magnetic fields is typically explained by a balance between magnetic exchange interaction, DM interaction, and magnetocrystalline anisotropy, it is clear that a change of any of these parameters upon application of an electric field will modify the energy landscape.

We have shown that while an external magnetic field adjusts the relative energy levels between skyrmion and ferromagnet globally, the same can be achieved locally by an electric field, see sketch in Fig. 4c, enabling full control over individual magnetic states. For our system we find that an electric field of 1 V/nm corresponds to a magnetic field of about 40 mT, Fig. 4d. This kind of electric field driven skyrmion switching may be superior over current-induced switching and demonstrates the feasibility of energy-efficient skyrmion devices.

References:

1. E. Y. Tsymbal, Spintronics: Electric toggling of magnets. *Nature Mater.* **11**, 12-13 (2012).

2. F. Matsukura, Y. Tokura, H. Ohno, Control of magnetism by electric fields. *Nature Nanotechnol.* **10**, 209-220 (2015).

3. S. Zhang, Spin-Dependent Surface Screening in Ferromagnets and Magnetic Tunnel Junctions. *Phys. Rev. Lett.* **83**, 640-643 (1999).

4. M. Weisheit, S. Fähler, A. Marty, Y.Souche, C. Poinsignon, D. Givord, Electric Field-Induced Modification of Magnetism in Thin-Film Ferromagnets. *Science* **315**, 349-351 (2007).

5. O. O. Brovko, P. Ruiz-Díaz, T.R. Dasa, V.S. Stepanyuk, Controlling magnetism on metal surfaces with non-magnetic means: electric fields and surface charging. *J. Phys.: Condens. Matter* **26**, 093001 (2014).

6. C.-G. Duan, J. P. Velev, R. F. Sabirianov, Z. Zhu, J. Chu, S. S. Jaswal, E. Y. Tsymbal, Surface Magnetoelectric Effect in Ferromagnetic Metal Films. *Phys. Rev. Lett.* **101**, 137201 (2008).

7. K. Nakamura, R. Shimabukuro, Y. Fujiwara, T. Akiyama, T. Ito, A. J. Freeman, Giant Modification of the Magnetocrystalline Anisotropy in Transition-Metal Monolayers by an External Electric Field. *Phys. Rev. Lett.* **102**, 187201 (2009).

8. M. Oba, K. Nakamura, T. Akiyama, T. Ito, M. Weinert, A. J. Freeman, Electric-Field-Induced Modification of the Magnon Energy, Exchange Interaction, and Curie Temperature of Transition-Metal Thin Films. *Phys. Rev. Lett.* **114**, 107202 (2015).

9. W. Eerenstein, N. D. Mathur, J. F. Scott, Multiferroic and magnetoelectric materials. *Nature* **442**, 759-765 (2006).

10. L. Gerhard, T. Yamada, T. Balashov, A. Takacs, R. Wesselink, M. Däne, M. Fechner, S. Ostanin, A. Ernst, I. Mertig, W. Wulfhekel, Magnetoelectric coupling at metal surfaces. *Nature Nanotechnol.* **5**, 792 (2010).

11. T. Maruyama, Y. Shiota, T. Nozaki, K. Ohta, N. Toda, M. Mizuguchi, A. A. Tulapurkar, T. Shinjo, M. Shiraishi, S. Mizukami, Y. Ando, Y. Suzuki, Large voltage-induced magnetic anisotropy change in a few atomic layers of iron. *Nature Nanotechnol.* **4**, 158 (2009).

12. Y. Shiota, T. Nozaki, F. Bonell, S. Murakami, T. Shinjo, Y. Suzuki, Induction of coherent magnetization switching in a few atomic layers of FeCo using voltage pulses. *Nature Mater.* **11**, 39-43 (2012).

13. W.-G. Wang, M. Li, S. Hageman, C. L. Chien, Electric-field-assisted switching in magnetic tunnel junctions. *Nature Mater.* **11**, 64-68 (2012).

14. U. Bauer, L. Yao, A. J. Tan, P. Agrawal, S. Emori, H. L. Tuller, S. van Dijken, G.S.D. Beach, Magneto-ionic control of interfacial magnetism. *Nature Mater.* **14**, 174-181 (2015).

15. N. Nagaosa, Y. Tokura, Topological properties and dynamics of magnetic skyrmions. *Nature Nanotechnol.* **8**, 899 (2013).

16. K. von Bergmann, A. Kubetzka, O. Pietzsch, R. Wiesendanger, Interface-induced chiral domain walls, spin spirals and skyrmions revealed by spin-polarized scanning tunneling microscopy. *J. Phys.: Condens. Matter* **26**, 394002 (2014).

17. N. Romming, C. Hanneken, M. Menzel, J. E. Bickel, B. Wolter, K. von Bergmann, A. Kubetzka, R. Wiesendanger, Writing and Deleting Single Magnetic Skyrmions. *Science* **341**, 636-639 (2013).

18. F. Jonietz, S. Muehlbauer, C. Pfleiderer, A. Neubauer, W. Muenzer, A. Bauer, T. Adams, R. Georgii, P. Boeni, R. A. Duine, K. Everschor, M. Garst, A. Rosch, Spin Transfer Torques in MnSi at Ultralow Current Densities. *Science* **330**, 1648-1651 (2010).

19. J. Sampaio, V. Cros, S. Rohart, A. Thiaville, A. Fert, Nucleation, stability and current-induced motion of isolated magnetic skyrmions in nanostructures. *Nature Nanotechnol.* **8**, 839 (2013).

20. A. Fert, V. Cros, J. Sampaio, Skyrmions on the track. *Nature Nanotechnol.* **8**, 152 (2013).

21. S. S. P. Parkin, M. Hayashi, L. Thomas, Magnetic Domain-Wall Racetrack Memory. *Science* **320**, 190-194 (2008).

22. P.-J. Hsu, A. Finco, L. Schmidt, A. Kubetzka, K. von Bergmann, R. Wiesendanger, Guiding spin spirals by local uniaxial strain relief. *Phys. Rev. Lett.* **116**, 017201 (2016).

23. S. Heinze, K. von Bergmann, M. Menzel, J. Brede, A. Kubetzka, R. Wiesendanger, G. Bihlmayer, S. Blügel, Spontaneous atomic-scale magnetic skyrmion lattice in two dimensions. *Nature Phys.* **7**, 713-718 (2011).

24. S. Heinze, Simulation of spin-polarized scanning tunneling microscopy images of nanoscale non-collinear magnetic structures. *Appl. Phys. A* **85**, 407-414 (2005).

25. C. Hanneken, F. Otte, A. Kubetzka, B. Dupé, N. Romming, K. von Bergmann, R. Wiesendanger, S. Heinze, Electrical detection of magnetic skyrmions by tunnelling non-collinear magnetoresistance. *Nature Nanotechnol.* **10**, 1039 (2015).

26. A. Sonntag, J. Hermenau, A. Schlenhoff, J. Friedlein, S. Krause, R. Wiesendanger, Electric-Field-Induced Magnetic Anisotropy in a Nanomagnet Investigated on the Atomic Scale. *Phys. Rev. Lett.* **112,** 017204 (2014).

**Acknowledgements** We thank B. Dupé, S. Heinze, A. Sonntag, and J. Hagemeister for discussions. Financial support from the Deutsche Forschungsgemeinschaft via SFB 668-A8, from the European Union's Horizon 2020 research and innovation programme under grant agreement No. 665095, and from the Hamburgische Stiftung für Wissenschaften, Entwicklung und Kultur Helmut und Hannelore Greve is gratefully acknowledged.

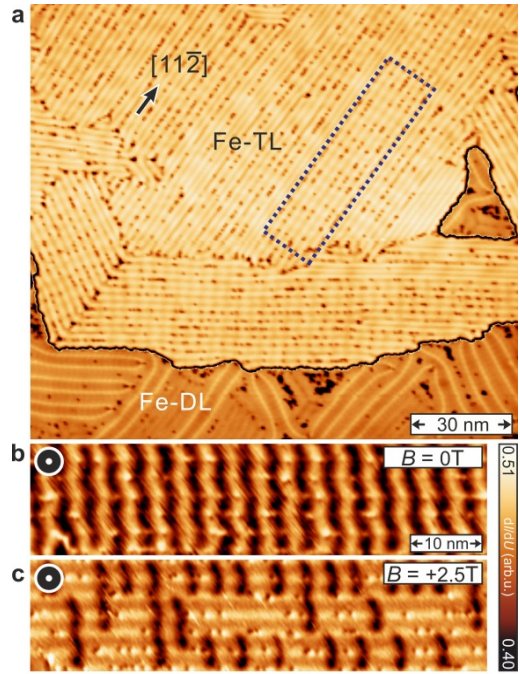

**Fig. 1. Field-dependent magnetic states of the Fe triple layer on Ir(111).** (**a**) SP-STM constant-current image of about 2.7 atomic layers of Fe on Ir(111) showing the sample topography. The contrast has been adjusted separately for the different terraces. Both the Fe triple layer (TL) and the Fe double layer (DL) are reconstructed due to uniaxial strain relief (see Fig. S1 for a structure model). (**b**) Corresponding spin-resolved d$I$/d$U$ map of the area indicated by the dashed rectangle in (a); the tip magnetization direction $m_t$ is indicated. At $B = 0$ T the magnetic ground state of the Fe-TL is a spin spiral with zigzag wavefronts and here the periodicity is about 3.8 nm. (**c**) Spin-resolved d$I$/d$U$ map of the same area in an external out-of-plane magnetic field of $B = +2.5$ T, the zero field spin spiral has broken up into individual magnetic objects (measurement parameters for all: $U = -0.7$ V, $I = 1$ nA, $T = 7.8$ K, Cr bulk tip which is sensitive to the out-of-plane sample magnetization).

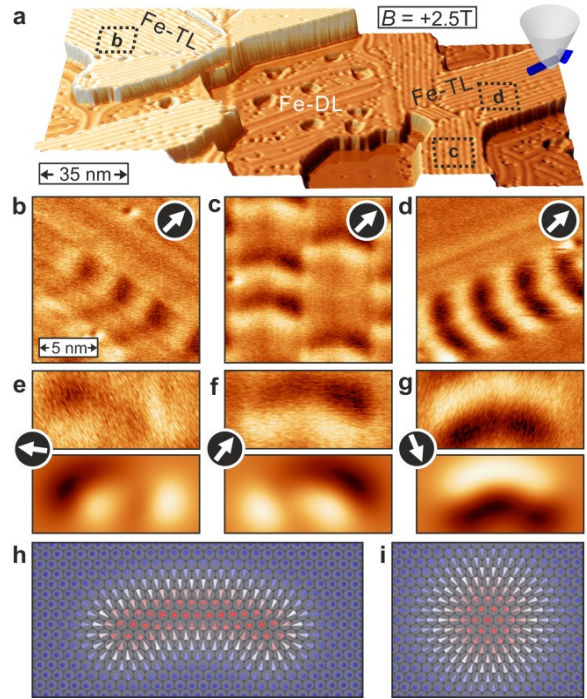

**Fig. 2. Spin structure of the magnetic objects.** (**a**) Perspective view of an SP-STM constant-current image of about 2.5 atomic layers of Fe on Ir(111) measured at $B = +2.5$ T; from the data we infer a tip magnetization as sketched. (**b-d**) Spin-resolved d$I$/d$U$ maps of each of the three possible rotational domains of the reconstructed Fe triple layer at $B = +2.5$ T, areas indicated in (a) by dashed squares; the tip magnetization direction is indicated by the arrows. (**e-g**) Spin-resolved d$I$/d$U$ maps (top) of one representative magnetic object of each rotational domain, rotated to have the dislocation lines vertical (image area is 7.5 nm × 3.5 nm), and SP-STM simulations (bottom) of the spin structure displayed in (h); the tip magnetization direction is indicated by the arrows. (**h**) The spin structure of the magnetic objects of the Fe triple layer, derived from the in-plane magnetic contrasts; image size same as in (e)-(g). (**i**) Spin structure of an axially symmetric magnetic skyrmion for comparison. (Measurement parameters for all: $U = -0.7$ V, $I = 1$ nA, $T = 7.8$ K, Cr bulk tip).

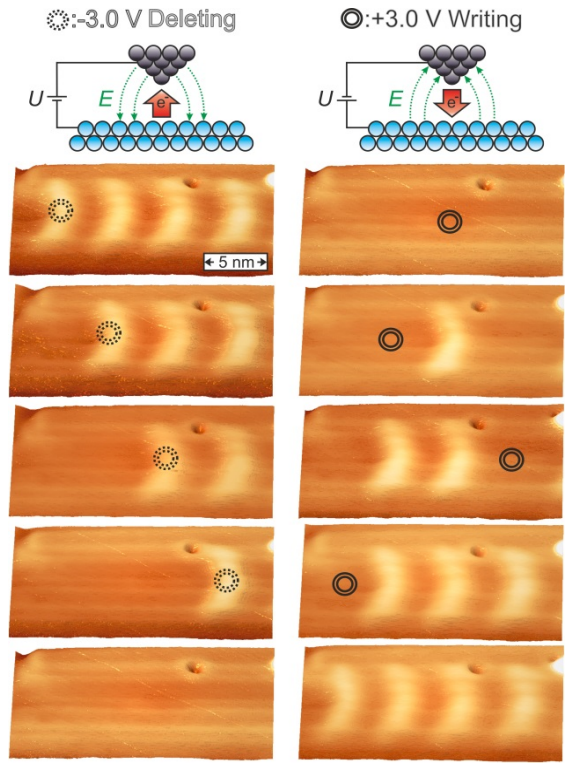

**Fig. 3. Writing and deleting of magnetic skyrmions.** Sketch of the experimental setup for writing and deleting (top). Perspective views of subsequent SP-STM constant-current images of the same Fe triple layer area (Cr bulk tip, $U$ = +0.3 V, $I$ = 0.5 nA, $T$ = 7.8 K, $B$ = +2.5 T). In between images voltage ramps up to $U$ = -3 V (+3 V) have been performed at the positions of the dashed (solid) circles, which reproducibly resulted in deleting (writing) of an individual magnetic skyrmion.

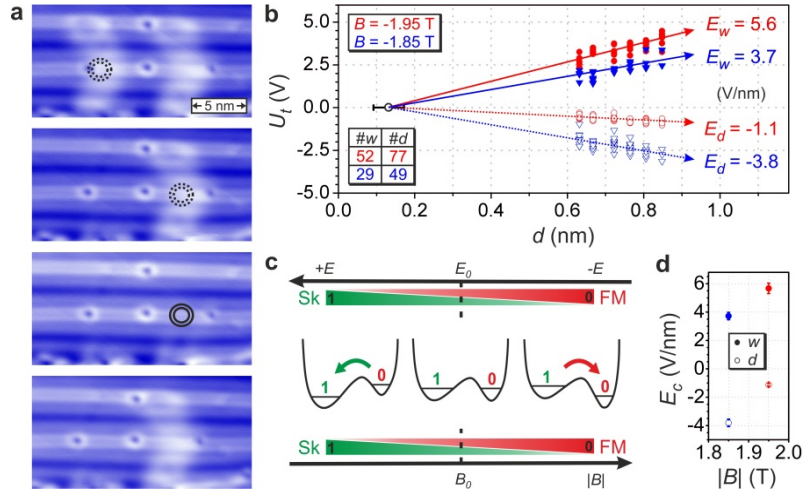

**Fig. 4. Electric field driven switching of magnetic skyrmions.** (**a**) Constant-current images of a measurement series with a spin-averaging W tip (measurement parameters: $U = +0.2$ V, $I = 1$ nA, $T = 7.8$ K, $B = -1.85$ T); again the deleting and writing is possible with voltages of -4 V (dashed circle) and +4 V (solid circle), respectively. (**b**) Threshold voltages $U_t$ for writing and deleting as a function of tip-sample distance $d$ for two different applied magnetic fields (see Figs. S2, S3 for the determination of the absolute tip-sample distance and the threshold voltage, respectively); #$w$,#$d$ indicate the number of data points; the success rate for switching was more than 95%. The linear dependence demonstrates the role of a critical electric field for writing and deleting, and the corresponding values $E_w$ and $E_d$ are given. (**c**) The sketches of the two state energy landscape illustrate that while a global tuning with the external magnetic field adjusts the relative energy levels between skyrmion and ferromagnet, a locally applied electric field can be used at the same time to favor one or the other state. (**d**) The critical electric fields with statistical error bars extracted from (b) plotted as a function of external magnetic field; we find that 1 V/nm corresponds to about 40 mT.

# Supplementary Materials for

# Electric field driven switching of individual magnetic skyrmions

Pin-Jui Hsu, André Kubetzka, Aurore Finco, Niklas Romming,
Kirsten von Bergmann*, Roland Wiesendanger

*Department of Physics, University of Hamburg, 20355 Hamburg, Germany*

*Corresponding author. E-mail: kbergman@physnet.uni-hamburg.de

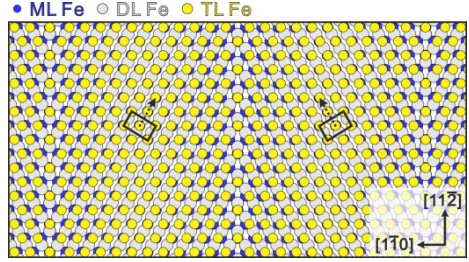

**Figure S1. Structure model and implications for the shape of magnetic states.** For the growth of Fe on Ir(111) a transition from pseudomorphic atom arrangement for the Fe monolayer (ML) to more bcc(110)-like Fe areas for higher layers is expected. A structure model that is in agreement with the experimentally observed symmetries is shown: the ML-Fe grows pseudomorphic to Ir(111) in fcc stacking, and for higher layers a uniaxial compression along the closed-packed atomic row leads to dislocation lines running along [11-2] (*22*). Locally the atomic arrangement varies periodically from fcc via bcc-like to hcp and further via bcc-like to fcc; two bcc(110) unit cells are indicated by rectangles. A similar coupling of the magnetic state to the local atom arrangement as presented for the Fe triple layer (TL) in the main text has also been found for the Fe double layer (DL) on Ir(111) (*22*). It can be pictured as a local alignment of the spin spiral propagation vectors $q$ with the bcc[001] direction (dashed arrows), which results in zigzag spin spiral wavefronts. In an applied magnetic field it is observed that the magnetic objects of the Fe-TL are always centered at the same type of hollow-site dislocation line, resulting in the bean-like shape with all beans in one structural domain pointing in the same direction, and a rotation of the bean-like shape by 120° between different rotational domains.

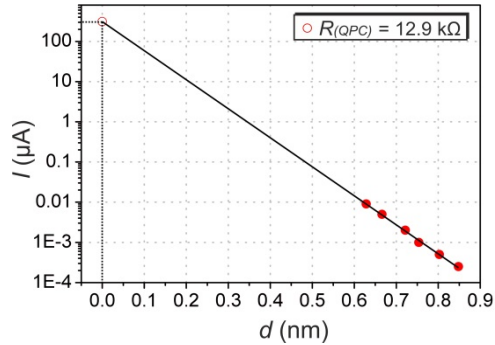

**Figure S2. Estimation of the absolute tip-sample distance for the measurement of Fig. 4.** The graph shows the tunnel current versus the relative tip-sample distances corresponding to $U$ = +4 V and the setpoint current values of the dataset in Fig. 4b. The linear behavior in this semi-logarithmic plot is characteristic for the tunnel regime. An extrapolation to the point contact resistance of 12.9 kΩ (*10*) yields an absolute tip-sample distance of 764±17 pm for 1 nA.

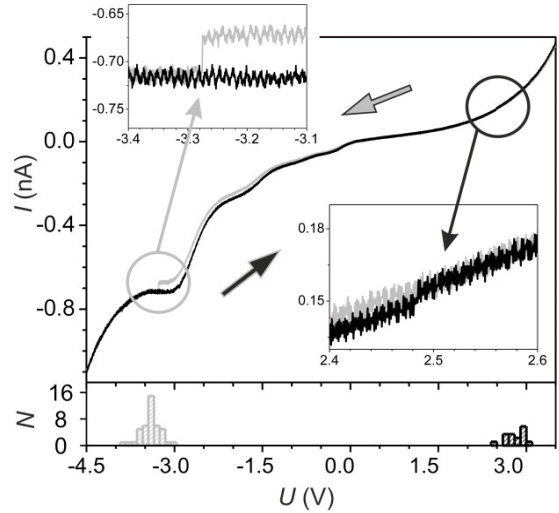

**Figure S3. Determination of the threshold voltages for switching.** $I(U)$ spectrum measured during the writing and deleting of one particular magnetic skyrmion. The jumps in the current mark the threshold voltages for writing at positive and for deleting at negative bias voltage; the insets show enlarged views of the switching events (stabilization parameters: $U = +4$ V, $I = 1$ nA; $T = 7.8$ K, $B = -1.75$ T). This measurement was repeated 42 times on the same sample position. The scattering of the respective threshold voltages is displayed at the bottom. Note that due to the small difference in the current signals for skyrmion and ferromagnet at positive bias voltage not all threshold voltages for writing could be extracted.

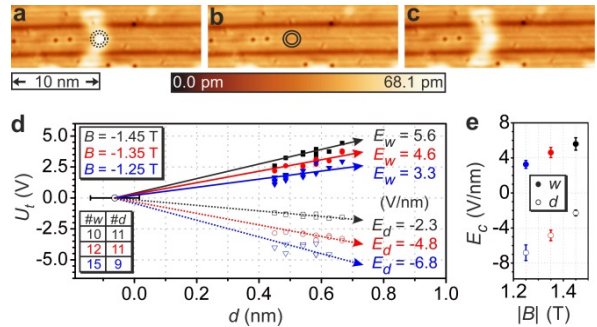

**Figure S4. Electric field driven skyrmion switching with a spin-polarized tip.** (**a-c**) Constant-current images of a measurement series with a spin-polarized Cr bulk tip, where one magnetic skyrmion has been deleted and rewritten with voltage ramps up to -4 V and +4 V, respectively (measurement parameters: $U = +0.2$ V, $I = 1$ nA, $T = 7.8$ K, $B = -1.45$ T). (**d**) Threshold voltages for writing and deleting for this measurement series at different tip-sample distances and applied magnetic fields; the linear behavior shows that also for the measurement with a spin-polarized tip the electric field dominates the switching mechanism. (**e**) The extracted critical electric field with statistical error bars versus the applied magnetic field. The difference to the results of Fig. 4b might be due to either a different distance between the dislocation lines at this sample area, a different surrounding, a different microtip that exhibits a different electrical field, or an additional contribution from spin-transfer torques. Note that the tip changed during this measurement series.